\documentclass[preprint,12pt]{elsarticle}
\usepackage{epsfig}
\DeclareGraphicsExtensions{.eps}

\usepackage{graphicx}
\usepackage{epstopdf}
 \usepackage{epsf}

\graphicspath{{../}}

\usepackage{amssymb}
\usepackage[top=2cm, bottom=2cm, left=2cm, right=2cm]{geometry}
\journal{SOLID STATE COMMUNICATIONS}
\begin{document}

\begin{frontmatter}



\title{Electronic structure and transport properties of CeNi$_9$In$_2$}

\author[label1]{R. Kurleto}
\author[label1]{P. Starowicz}
\ead{pawel.starowicz@uj.edu.pl}
\author[label2]{J. Goraus}
\author[label1]{S. Baran}
\author[label3]{Yu. Tyvanchuk}
\author[label3]{Ya.~M.~Kalychak}
\author[label1]{A.~Szytu{\l}a}

\address[label1]{Marian Smoluchowski Institute of Physics, Jagiellonian University, {\L}ojasiewicza 11, 30-348 Krakow, Poland}
\address[label2]{Institute of Physics, University of Silesia, Uniwersytecka 4, 40-007 Katowice, Poland}
\address[label3]{Department of Analytical Chemistry, Ivan Franko National University of Lviv, Kyryla and Mephodiya 6, 79005 Lviv, Ukraine}

\begin{abstract}
We investigated CeNi$_9$In$_2$ compound, which has been
considered as a mixed valence (MV) system. Electrical resistivity
vs. temperature variation was analysed in terms of the model
proposed by Freimuth for systems with unstable 4f shell. At low
temperature the resistivity dependence is consistent with a Fermi
liquid state with a contribution characteristic of
electron-phonon interaction. Ultraviolet photoemission
spectroscopy (UPS) studies of the valence band did not reveal a
Kondo peak down to 14~K. A difference of the spectra obtained
with photon energies of low and high photoionization cross
sections for Ce 4f electrons indicated that 4f states are located
mainly close to the Fermi energy. The peaks related to
f$_{5/2}$$^1$ and f$_{7/2}$$^1$ final states cannot be resolved
but form a plateau between -0.3 eV and the Fermi energy. X-ray
photoemission  spectroscopy (XPS) studies were realized for the
cerium 3d level. The analysis of XPS spectra within the
Gunnarsson-Sh\"{o}nhammer theory yielded a hybridization
parameter of 104 meV and non-integer f level occupation, being
close to 3. Calculations of partial densities of states were
realized by a full potential local orbital (FPLO) method. They
confirm that the valence band is dominated by Ni 3d states and
are in general agreement with the experiment except for the
behavior of f-electrons.
\end{abstract}

\begin{keyword}
A. Cerium intermetallics D. Electronic structure D. Transport
properties

\end{keyword}

\end{frontmatter}

%% \begin{linenumbers}
\section{INTRODUCTION}
\label{1} Various fascinating physical phenomena may arise when
electrons from conduction band hybridize with weakly localized
f-electrons \cite{a1}. The phases which are realized in the
system, methodized on a Doniach diagram \cite{b2}, depend on a
strength of the hybridization, determined by exchange integral or
equivalently by a location of 4f cerium atomic level with regard
to the conduction band. If the hybridization is weak, it may
allow for a long range magnetic ordering. A stronger
hybridization results in a Kondo state. Finally, in a case of a
very strong hybridization the f-electrons become delocalized and
form heavy fermions, mixed valence (MV) state or even participate
in a metallic valence band with lighter effective masses.

Crystal structure of CeNi$_9$In$_2$ (Fig. 1) belongs to the
YNi$_9$In$_2$ type with lattice constants a=b=8.2340(16) \AA~and
c=4.8310(9) \AA. The structure is characterized by large
coordination numbers (CN) for all the atoms~\cite{Bigun}. Namely,
we have 20 vertices coordination polyhedron (CP) for Ce atoms,
icosahedral CP with CN equal 12 for all Ni atoms and 15 vertices
CP for In. In this structure pairs of In atoms have very short
interatomic distances (2.639 \AA~ for CeNi$_9$In$_2$ as compared
to a double radius of In equal 3.252 \AA~\cite{Emsley}), therefore
a strong In-In interactions can be expected. Each In-In pair is
surrounded by 4 Ce and 19 Ni atoms.

Physical properties of CeNi$_9$In$_2$ have been reported before
\cite{Bigun, b17k, b18}. A specific heat of CeNi$_9$In$_2$ is
well described by the sum of electronic and lattice contributions
\cite{b18}. A Sommerfeld coefficient $\gamma$ equals 72
mJ/(mol$\cdot$K$^2$), while Debye temperature reaches a value of
139 K \cite{b18}. Previous studies revealed a concave shape of
electrical resistivity vs. temperature curve, which was supposed
to originate from MV state present in the system \cite{b17k,
Bigun}. Furthermore, an absence of magnetic ordering was reported
even at low temperatures. Magnetic susceptibility collected at a
high magnetic field (1 T) is nearly temperature independent
\cite{b17k}. On the other hand, magnetic susceptibility at low
magnetic field (50 Oe) reveals a broad maximum typical of MV
state \cite{b18}. However, further studies of the magnetic
susceptibility (unpublished) revealed that its values are too
high to follow the model proposed by Sales and Wohlleben
\cite{b6} for MV state. In fact, a presence of ferromagnetic
impurities in a specimen may be a source of the enlarged value of
susceptibility.

Electronic structure of CeNi$_9$In$_2$ was probed by means of XPS
\cite{b18} at room temperature. The authors assigned mainly Ni 3d
character to the valence band.  In order to describe mixing
between localized and conduction electrons
Gunnarsson-Sh\"{o}nhammer theory was employed \cite{b15, b15a}.
It yielded the value of hybridization equal to 105 meV and the
mean occupation of 4f level equal 0.84. This non-integer value of
the valence supports the presence of MV state in the
CeNi$_9$In$_2$ intermetallic.

In this paper we present studies of transport properties and
electronic structure of CeNi$_9$In$_2$, which were realized by
electrical resistivity measurements, ultraviolet photoemission
spectroscopy (UPS), x-ray photoemission spectroscopy (XPS) and
density functional theory calculations. The obtained results are
compatible with MV state, which has been proposed for this
compound.

\begin{figure}
\begin{center}
\includegraphics[width=3.4in]{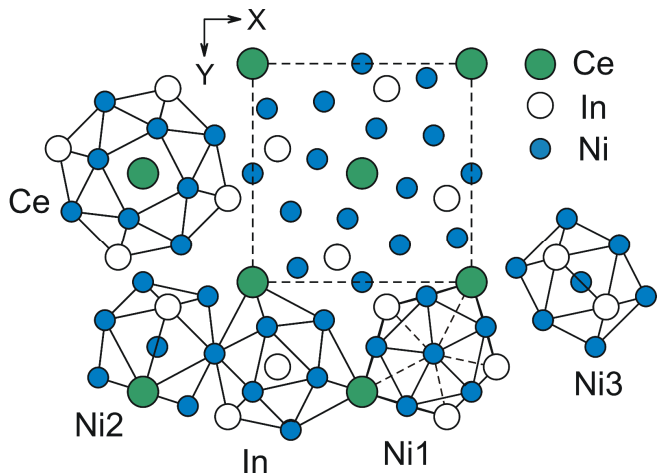}
\caption{(Color online) Crystal structure of CeNi$_{9}$In$_{2}$;
projection onto ab (xy) plane and coordination polyhedra.}
\end{center}
\end{figure}

\section{EXPERIMENTAL}
\label{2} A synthesis and characterization of polycrystalline
CeNi$_9$In$_2$ samples were described elsewhere \cite{b18}.
Electrical resistivity was measured with physical property
measurement system (PPMS, Quantum Design) with the application of
a four probe method. The measurements were realized in the
temperature range of 2 - 300 K.

The samples were studied by XPS and UPS using VG Scienta R4000
photoelectron spectrometer. UPS measurements were conducted with
He~I (21.2 eV) and He~II (40.8 eV) radiation and energy
resolution of 25 meV at the temperature of 14, 100, 200 and 300
K. XPS studies were realized by MgK$_\alpha$ (1254 eV) radiation
without a monochromator at 295 and 14 K. Sample surface was
prepared by scratching with a diamond file in ultra-high vacuum
of 5$\cdot$10$^{-10}$ mbar. The base pressure during the
experiment was 5$\cdot$10$^{-11}$ mbar. Energy calibration was
based on the measurements of Au 4f states and the Fermi energy
($E_{F}$) on Cu for XPS and UPS, respectively.

Density of states (DOS) for the CeNi$_9$In$_2$ compound was
calculated using full potential local orbital (FPLO) code in the
scalar relativistic version \cite{a2}. For the calculations the
Perdew -- Wang exchange - correlation potential was applied
\cite{a3, a3aa}. The local spin density approximation LSDA+U in
around mean field scheme was used in order to involve additional
correlations \cite{a4}. The value of Coulomb repulsion U in 4f
cerium shell was set equal to 6 eV as it is typically assumed
\cite{b3}.

\section{RESULTS AND DISCUSSION}
\label{3}
\subsection{ELECTRICAL RESISTIVITY}
Electrical resistivity presented in this paper varies with
temperature in the way consistent with the previous study
\cite{b17k}. A typical metallic shape with a small concavity
(Fig. 2) is clearly seen. It resembles the results obtained for
classical f-electron compounds with intermediate valence
\cite{a5}. Therefore, for higher temperatures (35 K - 300 K) we
have fitted the model proposed by Freimuth \cite{a6} for the
systems with unstable 4f shells. This model takes into account
the scattering between s, d electrons from conduction band and
f-electrons forming a narrow band. The electron phonon
interaction is represented by a linear term. The formula for the
resistivity reads:
\begin{equation}\rho(T) = \rho_0 + a\cdot T + b\cdot J_{sf}^2 \cdot
\frac{W(T)}{T_0^2+W(T)^2} ,\end{equation} where $W(T)=
T_{sf}\cdot e^{(-T_{sf}/T)}$ is effective energy width for
scattering. As fit results we obtained residual resistivity
$\rho_0$ = 7.38$\cdot$10$^{-7}$ $\Omega$$\cdot$m, a coefficient
in the phonon term $a$ = 1.09$\cdot$10$^{-9}$ $\Omega$$\cdot$m/K,
average f-electron position with respect to the Fermi energy
$T_0$ = 355 K, temperature characteristic of the fluctuations
between 4f$^0$ and 4f$^1$ states (fluctuation temperature)
$T_{sf}$= 157 K and the parameter related to the strength of
hybridization between s,d- and f-electrons $b\cdot J^{2}_{sf}$~=~45.2$\cdot$~10$^{-5}$~$\Omega$$\cdot$m$\cdot$K, where $J_{sf}$ is
the overlap of s, d- and f-electron wave functions, while b is
defined as $b=m\cdot k_B/ne^2\hbar$ \cite{a6}. The values of
$T_0$,  $T_{sf}$ and $b \cdot J^{2}_{sf}$ are relatively high when
compared to previous applications of the Freimuth's model
\cite{a6, a7} but all of them are within the same order of
magnitude. Also the Gr\"{u}neisen - Bloch - Mott \cite{b20,
b20aa} formula was fitted to the resistivity data. However, the
fit yielded a weaker agreement with the experiment, what was
reflected in lower correlation coefficient - $R^2$.

At low temperatures, namely in the range of 2 - 38 K, we have
fitted the formula:
\begin{equation}\rho=\rho_0+A\cdot T^2+B\cdot
T^5, \end{equation} so as to describe the system in terms of the FL
theory. Indeed the term with $A$ coefficient ($A$~=~7.7$\cdot$10$^{-12}$ $\Omega$$\cdot$m/K$^2$) is dominating, which is
characteristic of FL but the term with $B$ ($B$~=~2.1$\cdot$10$^{-16}$ $\Omega$$\cdot$m/K$^5$) is also important,
signifying that the contribution originating from dissipation on
the lattice cannot be neglected. We have obtained a large value
of residual resistivity that equals 7.6$\cdot$10$^{-7}$
$\Omega$$\cdot$m, hence a residual resistivity ratio (RRR)
reaches the value of 1.8.

Another verification of FL behavior \cite{b16} comes from the
Kadowaki -- Woods (KW) ratio calculated as $A/\gamma^2$, where
$\gamma$ is a Sommerfeld electronic heat capacity coefficient
taken from Ref. \cite{b18}. The obtained KW ratio amounted to
1.5$\cdot$10$^{-9}$
$\Omega\cdot$m$\cdot$mol$^{2}$$\cdot$K$^{2}$/J$^{2}$. According
to the reference \cite{b21} KW ratio in heavy fermions is equal
to 10$^{-7}$
$\Omega\cdot$m$\cdot$mol$^{2}$$\cdot$K$^{2}$/J$^{2}$, while in
transition metals ought to reach the value of 4$\cdot$10$^{-9}$
$\Omega\cdot$m$\cdot$mol$^{2}\cdot$K$^{2}$/J$^{2}$. The value
estimated for CeNi$_9$In$_2$ is of the same order of magnitude as
for transition metals. The deviation from the exact value
corresponding to transition metals may be due to the fact that
real calculated KW ratio should take into account corrections for
material specific parameters \cite{b21}. The other possibility
could be a breakdown of FL theory. On the other hand, the
proximity of CeNi$_9$In$_2$ and transition metals KW ratio is not
startling for us, as the valence band is dominated by nickel 3d
electrons \cite{b18}, what is also discussed further in the text.

\begin{figure}
\begin{center}
\includegraphics[width=3.4in]{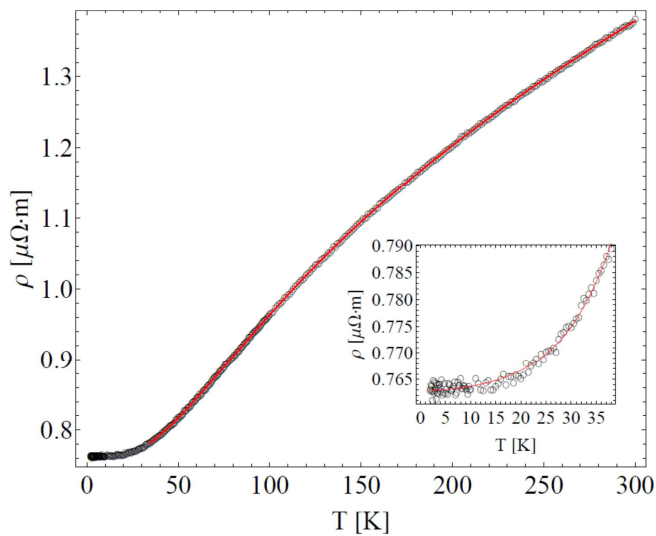}
\caption{(Color online) Electrical resistivity of
CeNi$_{9}$In$_{2}$ as a function of temperature. Experimental
data are marked with circles. The solid red line represents the
Freimuth model expressed by equation (1) fitted in the
temperature range 35 K to 300 K. The low temperature data (inset)
are fitted with equation (2).}
\end{center}
\end{figure}

\subsection{VALENCE BAND PHOTOEMISSION SPECTRA}

Manifestations of a Kondo effect and heavy fermion physics may be
visible in electronic structure, especially around $E_{F}$
\cite{a8, a9a}. So as to examine how many body interactions
impact on density of states we collected photoemission spectra
(UPS) of the valence band. The spectra recorded with the
application of He~I (21.2 eV) and He~II (40.8 eV) radiation at the
temperature T=14~K are presented together with their difference
in Fig. 3. Each spectrum is normalized to area in the region
from -3 eV to $E_{F}$.

The first important fact is that an intensive coherent peak near
$E_{F}$ called a Kondo peak was not found. Such a peak,
representing the final state with Ce f$_{5/2}^1$ configuration
was also not resolved in the spectra normalized by the
Fermi-Dirac distribution (not shown). UPS spectra obtained for
the valence band at different photon energies have generally
similar shape (not shown). He~I spectrum (Fig. 3) has two
maxima: the first, more intensive, at about -0.67 eV and the
second smaller and blurred at about -1.8 eV. Below the energy of
-3 eV one observes a contribution from inelastic background,
therefore it is not shown. The spectrum obtained with He~II
radiation also has two maxima: at  -0.6 eV and at about -1.8 eV.
The maxima in the He~II spectrum are better separated than those
in the case of He~I. The low energy peak for He~II is broader and
has increased spectral weight in a vicinity of $E_{F}$. This can
originate from Ce f electrons, which very often contribute to a
spectrum near $E_{F}$ with the f$^1$ final state \cite{a8}.

\begin{figure}
\begin{center}
\includegraphics[width=3.4in]{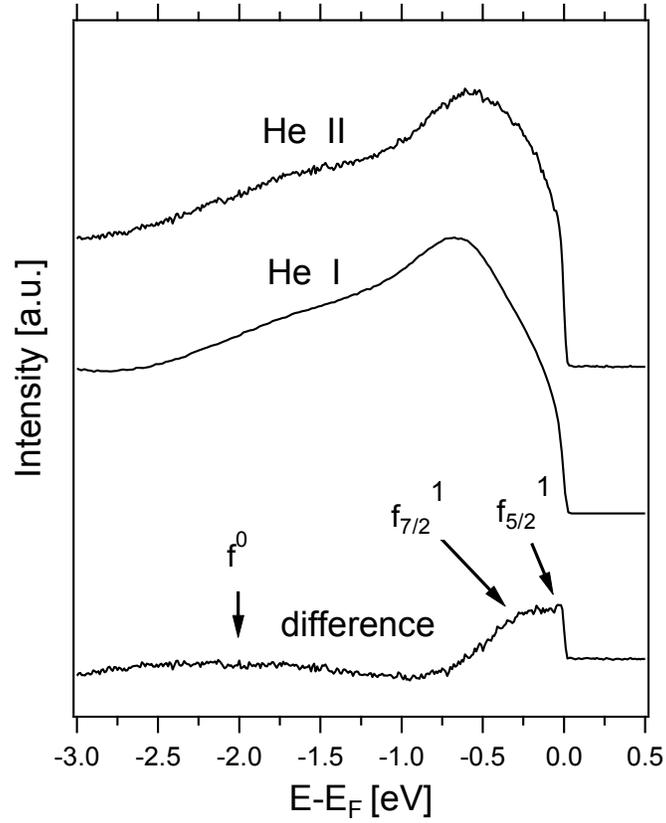}
\caption{Ultraviolet photoemission spectra obtained with He~II
(40.8 eV) and He~I (21.2 eV) radiation at the temperature of 14
K. The difference denotes a subtraction of He~I from He~II data.
Arrows indicate the energies of f$_{7/2}^{1}$ , f$_{5/2}^{1}$ and
f$^{0}$ final state peaks observed in cerium Kondo systems
\cite{a8,a11,a12}.}
\end{center}
\end{figure}

It was proved both theoretically \cite{a10} and experimentally
\cite{a8, a11} that Ce 4f electrons exhibit a higher
photoionization cross section for He~II radiation when compared
to He~I. The theoretical cross section is 3.25 times higher for
He~II \cite{a10} but the real increase of this factor according
to the experiment may be much larger as in many cases a sharp and
intensive Kondo peak is observed near the photon energy h$\nu$=
40 eV, while it vanishes at h$\nu$ = 20 to 25 eV \cite{a8, a11,
a12}. Therefore, we interpret the difference between He~II and
He~I spectra as originating from Ce 4f electrons (Fig. 3). Such a
difference (Fig. 3) shows low intensity in the range from -3.0 eV
to -0.6 eV with a broad maximum centered at about -2.3 eV. For
low binding energy one observes a high intensity in the range of
-0.3 eV to E$_F$. The difference resembles a typical spectral
function obtained from the single impurity Anderson model for a
Kondo system \cite{a11} but does not show sharp peaks. A high
intensity near E$_F$ should accommodate both a f$_{5/2}^1$ peak
observed at $E_{F}$ and a spin orbit partner f$_{7/2}^1$ located
in the region of $\sim$ -0.28 meV \cite{a8, a12}. Despite a very
high energy resolution of our UPS study the peaks f$_{5/2}^1$ and
f$_{7/2}^1$ are not resolved but form a broad maximum or rather a
plateau between 0.3 eV and $E_{F}$. According to the calculations
based on Anderson model \cite{a13, a14, a15} the Kondo peak is
broadened in the case of non-integer f-level occupation, which
takes place in MV state. Such a situation is reflected in our
spectra. Kondo peak would also be absent or much broadened, if
the spectra are collected at the temperature considerably higher
than the Kondo temperature ($T_{K}$). However, we do not favor
such a hypothesis, as a very low value of $T_{K}$ (much below 14
K) is not in agreement with the valence fluctuation scenario in
CeNi$_9$In$_2$. A broad peak in the region of -2.3 eV of the
difference spectrum may be attributed to the f$^0$ final state
\cite{a8,a12}. UPS studies were conducted in a series of
temperatures between T=14 K and room temperature but no clear
differences have been found. It is known from theoretical
calculations (discussed further in the text) that the valence
band of CeNi$_9$In$_2$ is dominated by Ni 3d and Ce 4f electrons.
According to what has been written, the spectrum obtained by He~I
may be treated as originating mainly from the Ni 3d electrons.

\subsection{XPS ON 3D CERIUM LEVELS}

Holes, which are created in 3d or 4d cerium levels as a
consequence of a photoionization, strongly interact with 4f
electrons. This effect gives us an opportunity to characterize a
hybridization between strongly correlated f electrons and a
conduction band. The aforementioned interaction between a hole and
electrons affects the peaks related to f$^0$, f$^1$ and f$^2$
final states. Thus, each peak in the d-electron doublet is split
into three components. While the 4d level spectrum has a complex
structure, peaks in the 3d cerium spectrum can be resolved with
the application of Doniach-\v{S}unji\'{c} theory \cite{a16}.

The XPS spectrum of Ce 3d level for CeNi$_9$In$_2$ is shown in the
Fig 4. Features observed at low binding energy side originate
from nickel 2p states. These are 2p$_{1/2}$ nickel main peak at
about 869.70 eV and satellites at 873.40 eV and 875.70 eV
\cite{tut}. The spectrum of 3d cerium consists of spin-orbit
partners separated by 18.5 eV. Each spin-orbit partner has three
distinguishable components. The main peak corresponding to the
f$^1$ final state is accompanied by f$^2$ and f$^0$ satellites.
Locations of particular peaks are given in table 1. Non-zero
intensity of f$^2$ satellite is a corollary of screening and
corroborates that f states hybridize with a Fermi sea. In order to
obtain values of hybridization strength and f level filling the
Gunnarsson-Sh\"{o}nhammer theory was employed \cite{b15, b15a}.
We have calculated the following intensity ratios:
$r_0=I(f^0)/[I(f^0)+I(f^1)+I(f^2)]$ and
$r=I(f^2)/[I(f^1)+I(f^2)]$. Subsequently, we have used $r_0$ in
the formula: $n_f=1-r_0$, and then we have deduced the value of
the hybridization from $r$, according to the reference \cite{b15,
b15a}. We obtained the hybridization $\Delta$=104 meV and f-level
occupation $n_f$ = 0.94. The parameters $\Delta$ and $n_f$
slightly differ from that determined before \cite{b18}. Actually,
different circumstances should be counted, like differences in
sample preparation or in approach to spectra fitting. Previously
\cite{b18} the samples were fractured, while in this study
surface was cleaned using a diamond file. XPS spectra collected
at 295 K and 20 K do not exhibit any considerable difference.

\begin{figure}
\begin{center}
\includegraphics[width=3.4in]{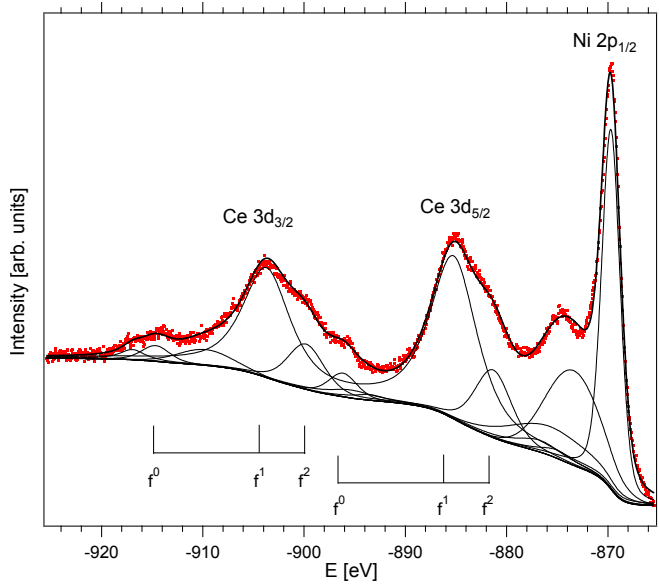}
\caption{(Color online) X-ray photoemission spectra recorded by Mg
K$_{\alpha}$ radiation at room temperature. Ce 3d$_{5/2}$ and Ce
3d$_{7/2}$ levels are fitted with three pairs of peaks related to
f$^{0}$, f$^{1}$ and f$^{2}$ final states. Spin-orbit splitting
equals 18.5 eV.}
\end{center}
\end{figure}

\label{tab}

\begin{center}
\begin{table}
\caption{Binding energies ($E_{B}$) and intensities (I) of the
peaks fitted to XPS spectrum of Ce 3d$_{5/2}$ level. f$^{0}$,
f$^{1}$ and f$^{2}$ represent configurations with 0, 1 and 2
electrons, respectively which occupy f shell in the final state.
The binding energies for Ce 3d$_{3/2}$ peaks are shifted by the
spin-orbit splitting of 18.5 eV and their relative intensity
ratios are the same as for 3d$_{5/2}$.}
 \vspace{0.3in}
\centering
\begin{tabular}{ c c c c c c l}
\hline
\hline
&3d$_{5/2}$&\\
\hline
%\hline
f$^0$&f$^1$&f$^2$&\\
896.16&885.22&881.32&$E_{B}$ [eV]\\
132&1758&485&$I$ [arb. u.]\\
\hline
\hline
\\
\end{tabular}
\end{table}
\end{center}

\subsection{THEORETICAL CALCULATIONS}

In order to provide a comparison between the experimental results
and theory, we have performed FPLO calculations of DOS. Here we
present DOS obtained for non-correlated f level (U=0) and for f
electrons interacting with effective Coulomb repulsion that equals U=6
eV (Fig. 5). As expected, the spectral density for U=0 is roughly
spin independent. According to the calculations, the total
density of states in the valence band region is mainly built up
of nickel 3d states and cerium 4f states. While the former cover
the whole valence band, the latter form a narrow peak next to
$E_{F}$. Indium gives a small contribution, mainly from 5p and 5s
states. While 5p partial DOS reaches a reasonable value in the
range from about -6 to 2 eV, 5s states give contribution for
energy below -6 eV or above $E_F$ energy. In the case of U=~6eV a
situation is different. Total DOS depends on spin, because of a
broken symmetry in cerium partial DOS. The contribution from
indium and nickel atoms is the same as in U=0 case. 4f partial
DOS, which only plays a role for cerium atoms, forms two peaks
for each spin. Peaks for up spin are located at -4 eV and 2 eV,
while those for down spin at about 1.7 and 2.5 eV. DOS calculated
for Ni 3d states is in a reasonable agreement with the
experiment, while it is not a surprise that DOS obtained within
the used theoretical approach for Ce 4f electrons does not model
the real spectral function. The Sommerfeld coefficient was
calculated using the obtained theoretically value of DOS at
$E_F$. It amounts to 40.8 mJ/(mol$\cdot$K$^2$) and 33.5
mJ/(mol$\cdot$K$^2$) for the cases with U=0 and 6 eV,
respectively. Thus, the experimental value of 72
mJ/(mol$\cdot$K$^2$) \cite{b18} indicates the presence of
correlations in the system.

\begin{figure}
\begin{center}
\includegraphics[width=6.8in]{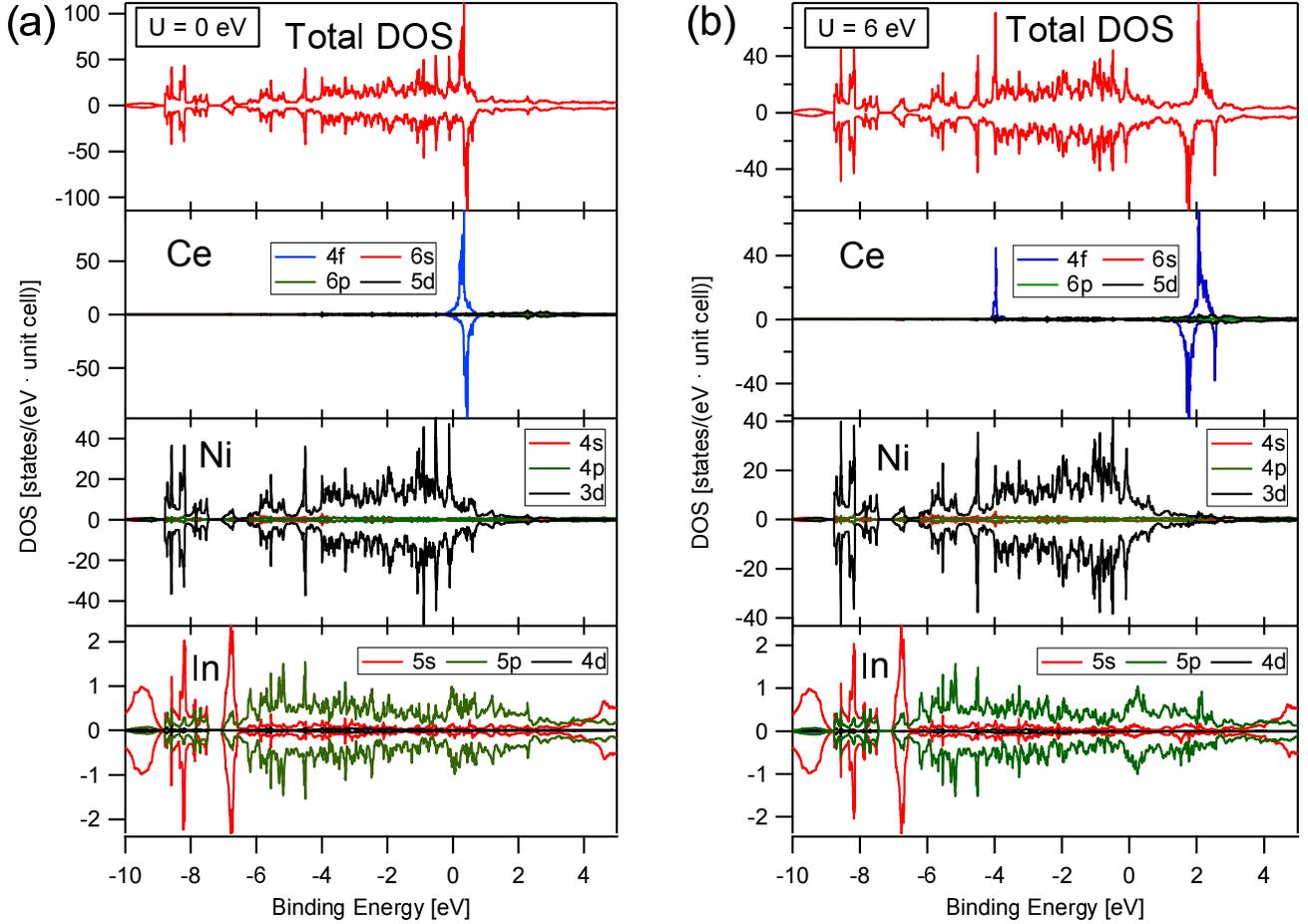}
\caption{(Color online) Total density of states (DOS) and partial
DOSes obtained theoretically by means of FPLO calculations with
the 4f electron correlation parameter U$=0$~eV (a) and U$=6$~eV
(b). Positive and negative values of DOS represent opposite spin
directions.}
\end{center}
\end{figure}
\section{SUMMARY}
\label{4}

We presented the studies of electrical resistivity, UPS, XPS and
theoretical calculations for MV compound CeNi$_9$In$_2$. The
electrical resistivity was analyzed within the model for unstable
valence 4f systems. Low temperature behavior of resistivity is
described by a quadratic function characteristic of a Fermi
liquid with a contribution originating from the dissipation on
the lattice. UPS studies did not reveal a Kondo peak down to 14
K. f-electron spectral function was obtained by a subtraction of
He~I spectrum from the He~II one. The peaks for the f$_{5/2}$$^1$
and f$_{7/2}$$^1$ final states form a plateau between 0.3 eV and
$E_{F}$, what should be related to their broadening. This is a
symptom of MV state in CeNi$_9$In$_2$. According to band
structure calculations the valence band is dominated by Ni 3d
electrons. The analysis of Ce 3d level by XPS allowed to
determine the hybridization between conduction band and
f-electrons as well as f-level occupation parameters. The
obtained Ce valence is non-integer but close to 3.

\section{ACKNOWLEDGMENTS}
R.K. and P.S. acknowledge fruitful discussions with V. H. Tran.
This work has been supported by the Ministry of Science and
Higher Education in Poland within the Grant no. N N202 201 039.
J.G. acknowledges the financial support from the National Science
Centre (NCN), on the basis of Decision No.
DEC-2012/07/B/ST3/03027. The measurements were carried out with
the equipment purchased thanks to the European Regional
Development Fund in the framework of the Polish Innovation
Economy Operational Program (contract no.
POIG.02.01.00-12-023/08).
\vspace{0.2in}
%\newpage

%% \end{linenumbers}



 %% \bibliographystyle{elsarticle-num}
 %% \bibliography{a}


%\begin{thebibliography}{00}


%\bibitem{}

%\end{thebibliography}

%%\section{Tables}

\end{document}